\documentclass[amscd,amsmath,amssymb,verbatim, aps, prd, showpacs]{revtex4}
\usepackage{amsmath, amsthm, amssymb}

\usepackage{graphicx}
\usepackage[dvips]{hyperref}
\usepackage{textcomp}
\usepackage[OT1,T1]{fontenc}

\begin{document}

\title{CMB signature of a super-Hubble inhomogeneity in the gravitational
field enclosing the present Hubble volume}
\author{Kjell Tangen}
\email{kjell.tangen@dnv.com}
\affiliation{DNV, 1322 H{\o}vik, Norway}\label{I1}
\date{\today}
\begin{abstract}
Repeated studies of the cosmic microwave background (CMB) based
on data from the Wilkinson Microwave Anisotropy Probe have revealed
an apparent asymmetry in the distribution of temperature fluctuations
over the celestial sphere. The studies indicate that the amplitudes
of\ \ temperature fluctuations are higher in one hemisphere than
in the other. We consider whether this asymmetry could\ \ originate
from a large scale inhomogeneity in the gravitational field enclosing
the present Hubble volume. We examine what effect the presence of
an inhomogeneity in the gravitational field of size larger than
the present Hubble radius would have on the temperature distribution
of the CMB and start eliciting its observational signature in the
CMB power spectrum. The covariance function $\langle a_{\mathit{lm}}^{*}a_{l^{\prime
}m^{\prime }}\rangle $ contains, in addition to the diagonal $C_{l}$
entries of the conventional CMB temperature anisostropy power spectrum,
non-diagonal entries. We find that specific non-diagonal entries
of the covariance function are sensitive to the strength of the
inhomogeneity, while the diagonal $C_{l}$ entries are not. These
non-diagonal entries, which are not present in the case of a homogeneous
background geometry, are observational signatures of a large-scale
inhomogeneity in the background geometry of the universe. Furthermore,
we find that an inhomogeneity in the gravitational potential of
super-Hubble size would yield a power asymmetry in the CMB with
maximal asymmetry at an angle of 90$ \mbox{}^{\circ}$ to the CMB
dipole axis. The axis of the CMB power asymmetry was recently estimated
by Eriksen et. al. to be located at angles between 83$ \mbox{}^{\circ}$
and 96$ \mbox{}^{\circ}$ to the CMB dipole axis, which is consistent
with the prediction of our model. This implies that the location
of the observed power asymmetry in the CMB sky could be accounted
for by a large-scale inhomogeneity in the gravitational field enclosing
the present Hubble volume.
\end{abstract}
\maketitle

\section{Introduction}

Precision measurement and analysis of the cosmic microwave background
(CMB) over the past two decades has exposed an amazingly detailed
picture of the early universe, a picture in which not only the processes
and structure of the early universe is imprinted, but also the signature
of any subsequent gravitational processing of this structure. This
makes the CMB an extraordinarily rich and powerful tool for probing
and depicting physical processes in the early universe.

Hence, analysis of the CMB has uncovered evidence of how the observed
structure in the universe has emerged: In the primordial universe,
stochastic processes in the form of quantum fluctuations in primordial
fields drove fluctuations in the gravitational field which subsequently
gave rise to fluctuations in the matter and photon densities. These
primordial density fluctuations are today imprinted in the temperature
fluctuations in the CMB, and the properties of the primordial processes
are encoded in the statistical properties of the CMB. In order to
understand the primordial universe, it is therefore of utmost importance
to understand the statistical properties of the CMB and decode its
statistical information.

Full-sky analysis of data from the Wilkinson Microwave Anisotropy
Probe (WMAP) are in good agreement with the predictions of simple
inflationary models that prescribe a nearly scale free, Gaussian
primordial power spectrum and a universe that is flat and homogeneous
at large scales \cite{Dunkley:2008ie,Komatsu:2008hk}. The CMB spectrum
expected in such a universe is statistically isotropic, implying
that the expectation value of CMB observables are rotationally invariant.
There is, however, an apparent asymmetry in the power of CMB flucutations
between two different hemispheres \cite{Eriksen:2003db}\cite{Hansen:2004vq,Hansen:2006rj,Eriksen:2007pc}
of the CMB sky. Various explanations of the anomaly have been proposed.
New inflationary models \cite{Ackerman-2007,Erickcek-2008l3520E,Watanabe:2009ct}
may yield asymmetric power spectra, which may account for the CMB
power asymmetry. Other authors\ \ have invoked conventional physics,
such as non-linear gravitational effects of local inhomogeneities
\cite{Tomita:2005nu} , or dust-filled voids in the local universe
\cite{Inoue:2006fn}.

Here, we are taking a simplistic phenomenological perspective where
we consider whether the presence of a large scale inhomogeneity
in the gravitational field enclosing the present Hubble volume could
account for the power asymmetry\footnote{It should be noted that,
although the visible universe would be inhomogeneous at scales larger
than the present Hubble volume (super-Hubble scales),\ \ a homogeneous
universe at even larger scales is not precluded.}. At this stage,
we are not concerned with how the inhomogeneity came about, but
it is natural to assume that it has primordial origin. However,
independent of its origin, a gravitational inhomogeneity in the
form of a large scale perturbation to the gravitational potential
could possibly manifest itself through a visible asymmetry in the
CMB power spectrum. If successful, we may be able to capture the
essentials of the physical phenomena under study in a rather model
independent way, without excluding relevant models that may account
for the phenomena.\ \ \

If we were to detect large-scale asymmetries in the CMB temperature
distribution, the determination of their origin, whether they originate
from asymmetries in the primordial power spectrum or from large
scale inhomogeneities in the gravitational field, would in any case
provide us with valuable additional insight into the nature of the
processes in the primordial universe.\ \

Given an early inflationary universe, large scale fluctuations in
the gravitational field will directly depict primordial processes,
because as they leave the horizon early in the inflationary era,
large scale scalar perturbations to the gravitational field freeze.
Moreover, they would stay more or less constant until now, only
decaying slightly throughout the history of the universe \cite{Dodelson-2003}.

How could such large-scale perturbations to the gravitational field
- of size larger than the observable universe - be observed? This
is the first question that we would like to answer in this paper.
Under the assumption that our observable universe is immersed in
a scalar metric perturbation of super-Hubble size, the aim of this
paper is to determine what effect the presence of such a perturbation
would have on the CMB temperature and start to elicit its observational
signature in the CMB spectrum.

Given the puzzling evidence of a hemispherical power asymmetry in
the CMB \cite{Eriksen:2003db}, a natrual follow-up question is:
Could an hemispherical\ \ power asymmetry in the CMB be caused by
a large-scale inhomogeneity in the gravitational field of super-Hubble
size? Posing these questions, we make no assumptions about the origin
of his perturbation or any physical mechanism that might have caused
it. Our goal is simply to determine what effect the presence of
such a perturbation to the gravitational potential would have on
the observed CMB temperature and to elicit its observational signature
in the CMB spectrum.

Naively, one could perhaps attempt to answer these questions by
looking for an axis of maximal anisostropy in the CMB spectrum \cite{Eriksen:2003db}
and correlate the apparent CMB anisotropy with the expected anisotropy
from a super-Hubble perturbation to the gravitational potential.
There is, however, already such an axis of maximal anisotropy: the
CMB dipole, which originates from our motion relative to the CMB
rest frame and by far dominates the CMB anisotropy spectrum. We
will therefore have to look for the signature of a secondary anisotropy
axis in the residual temperature map after first removing the CMB
dipole.

This paper starts by computing the expected temperature anisotropy
from a super-Hubble perturbation to the gravitational potential,
as seen from the CMB rest frame. Then, we transform this result
to a moving frame of observation, before we factor out the best
fit dipole. Treating the resulting residual as a lowest order perturbation
to the temperature distribution expected from a standard, homogeneous
background cosmology, we can therefore factor the residual into
the temperature distribution of an arbitrary homogeneous background
cosmology. By doing a standard multipole expansion of the temperature
distribution, we are finally able to obtain the signature effect
of the super-Hubble perturbation on the CMB spectrum.\ \ \

\section{Super-Hubble perturbations}\label{XRef-Section-92561115}

Let us briefly review the large scale solutions to the first order
perturbation equations for a $\Lambda$CDM cosmology. More details
can be found in Appendix \ref{XRef-AppendixSection-9723255}. and
in \cite{Dodelson-2003}.\ \ In conformal Newtonian gauge, the space-time
metric takes the form
\begin{equation}
{\mathrm{ds}}^{2}=a^{2}( \eta ) \left( -\left( 1+2\Psi ( \eta ,x)
\right) {\mathrm{d\eta }}^{2}+\left( 1+2\Phi ( \eta ,x) \right)
\delta _{\mathit{ij}}{\mathrm{dx}}^{i}{\mathrm{dx}}^{j}\right) ,%
\label{XRef-Equation-9817284}
\end{equation}

assuming implicit summation of spatial indices. $\eta $ is the conformal
time, while $\Psi $ and $\Phi $ are the two scalar metric perturbations.
In this section, we will restrict our attention to the evolution
of the gravitational potentials $\Phi $ and $\Psi $, and will therefore
discard any variable without significant contribution to their evolution.

Before we continue, let us define what we mean by a super-Hubble
perturbation. As we primarily will work in the spatial domain, we
need a definition of super-Hubble perturbations in the spatial domain
that is compatible with the conventional definition in the frequency
domain. In the frequency domain, a super-Hubble perturbation is
a mode with wave length much larger than the comoving Hubble radius
$\mathcal{H}^{-1}$, where $\mathcal{H}\equiv \dot{a}/a$. A super-Hubble
mode with wave vector $k_{i}$ will therefore have $k^{2}<< \mathcal{H}^{2}$.
Similarly, for any perturbation variable $X$, we define a spatial
domain super-Hubble perturbation to be any spatial perturbation
with spatial derivatives that satisfy $|\partial _{i}X/X|\ \ <<
\mathcal{H}$ and $|\partial ^{2}X/X| << \mathcal{H}^{2}$. By applying
this definition to a Fourier mode $\sim e^{i k\cdot x}$, we see
that the definitions of spatial-domain and frequency domain super-Hubble
perturbations are consistent.

At super-Hubble scales, the temporal dependence of the perturbation
variables can be separated from the spatial dependence, allowing
us to write the potential as a product of a fixed spatial perturbation
and a time-dependent factor:
\begin{equation}
\Phi \simeq -\Psi \simeq \Phi ( \eta ) \Phi _{0}( x)
\end{equation}

where $\Phi _{0}( x) $ is the primordial spatial perturbation. Now,
since $\Phi $ is a super-Hubble perturbation, we have\ \ $\mathcal{H}^{-1}\partial
_{i}\Phi _{0}<< \Phi _{0}$. Consequently, within the present Hubble
volume, higher-order derivatives of $\Phi _{0}$ can be ignored,
and $\Phi _{0}$ may be approximated by its first order Taylor expansion
around the position $x_{0}$ of the observer:
\begin{equation}
\Phi _{0}( x) \simeq \Phi _{0}( x_{0}) +\partial _{i}\Phi _{0}(
x_{0}) \left( x^{i}-x_{0}^{i}\right) %
\label{XRef-Equation-9961914}
\end{equation}

The spatial coordinates $\{x^{i}\}$ are arbitrary, and we need to
fix them in order to proceed.\ \ Let us choose cartesian spatial
coordinates $x,y,z$ that make the potential, and therefore the metric
of eq. (\ref{XRef-Equation-9817284}) manifestly axially symmetric
to first order in\ \ $\partial _{i}\Phi _{0}( x_{0}) $. Align the
$z$ axis with the gradien $\nabla \Phi _{0}( x_{0}) $ and let the
origin be the observer's position. Then the linear expansion of
the potential in eq. (\ref{XRef-Equation-9961914}) takes the form\ \ \ \
\begin{equation}
\Phi _{0}( z) \simeq -k \left( z-z_{0}\right)
\end{equation}

where the constant $k\equiv {(\delta ^{\mathrm{ij}}\partial _{i}\Phi
_{0}( x_{0}) \partial _{j}\Phi _{0}( x_{0}) )}^{1/2}$ and $z_{0}$
is a constant. With this approximation, the metric of eq. (\ref{XRef-Equation-9817284})
now takes an axially symmetric form:\ \
\begin{equation}
{\mathrm{ds}}^{2}=a^{2}( \eta ) \left( -\left( 1+2\Psi ( \eta ,z)
\right) {\mathrm{d\eta }}^{2}+\left( 1+2\Phi ( \eta ,z) \right)
\left( {\mathrm{dx}}^{2}+{\mathrm{dy}}^{2}+{\mathrm{dz}}^{2}\right)
\right) %
\label{XRef-Equation-91865515}
\end{equation}

In the following, we will use the space-time metric of eq. (\ref{XRef-Equation-91865515})
as the background metric. This background can be treated as a perturbation
to the homogeneous background\ \ ${\mathrm{ds}}^{2}=a^{2}( \eta
) (-d\eta ^{2}+\delta _{\mathrm{ij}}dx^{i}dx^{j})$. To first order
in the perturbation variables,\ \ this pertubation evolves independently
of other perturbations, a result of the perturbation equations being
linear. We are therefore able to treat the effect of a super-Hubble
metric perturbation independently of other effects.\ \

\section{The effect of a super-Hubble metric perturbation on the
angular CMB temperature distribution }

In order to compute the CMB temperature perturbation arising from
a super-Hubble metric perturbation, we need to solve the equations
of motion for photons arriving from an arbitrary direction $(\theta
,\varphi )$. In the following, we will assume a space-time metric
on the form
\begin{equation}
{\mathrm{ds}}^{2}=a^{2}( \eta ) \left( -A( \eta ,z) {\mathrm{d\eta
}}^{2}+{A( \eta ,z) }^{-1}\left( {\mathrm{dx}}^{2}+{\mathrm{dy}}^{2}+{\mathrm{dz}}^{2}\right)
\right) ,%
\label{XRef-Equation-918173048}
\end{equation}

where $A( \eta ,z) \equiv 1+2\Psi ( \eta ,z) $, which is an idealization
of the general case, but reduces to eq. (\ref{XRef-Equation-91865515})
to first order in $\Phi $ in the case $\Psi =-\Phi $.

Once we have computed the trajectories of free-streaming photons
arriving from an arbitrary direction, we will be able to compute
the gravitational potential at the time of emission. Assuming CMB
photons were emitted simultaneously at a uniform temperature at
recombination, the current CMB temperature will primarily depend
on the gravitational potential at the time and place of emission
as well as any subsequent development of the gravitational potential
along the paths of the CMB photons.

\subsection{Computing null geodesics}

The first item on our agenda is to compute the trajectories of photons
moving in the inhomogeneous background defined by eq. (\ref{XRef-Equation-918173048}).
These trajectories are null geodesics, determined by the geodesic
equations. The geodesic equations are derived by minimizing the
action $S=\int d\lambda \frac{1}{2}g_{\mu \nu }\frac{dx^{\mu }}{\mathrm{d\lambda
}}\frac{{\mathrm{dx}}^{\nu }}{\mathrm{d\lambda }}$ for a massless
particle moving in the background of\ \ eq. (\ref{XRef-Equation-918173048}).\ \ Null
geodesics, as well as certain geometric quantities, like angles,
are invariant under conformal transformations. We therefore choose
to compute the null geodesics in a different conformal frame, using
the conformally related metric
\begin{equation}
d{\overline{s}}^{2}=-A( \eta ,z) {\mathrm{d\eta }}^{2}+{A( \eta
,z) }^{-1}\left( {\mathrm{dx}}^{2}+{\mathrm{dy}}^{2}+{\mathrm{dz}}^{2}\right)
\end{equation}

The reason for this is that the equations of motion simplify considerably
in this metric. Notice that, even if the null geodesics and certain
other quantities are invariant under conformal transformations,
most physical properties of the photons depend on the conformal
frame. We will therefore take care to use the physical spacetime
metric of eq. (\ref{XRef-Equation-918173048}) when computing conformally
variant physical properties of the photons, such as the photon energy.

 Two constants of motion, $v_{x}$ and $v_{y}$, arise because coordinates
$x$ and $y$ are cyclic coordinates of the action. The equations
of motion for coordinates $x$ and $y$ are, using overdots to denote
differentiation with respect to the affine parameter $\lambda $:
\begin{equation}
A^{-1}\dot{x}= v_{x}, A^{-1}\dot{y}= v_{y}
\end{equation}

Furthermore, the equation for the time coordinate $\eta $ is
\begin{equation}
\frac{d}{\mathrm{d\lambda }}\left( A \dot{\eta }\right) -\frac{\partial
A}{\partial \eta }{\dot{\eta }}^{2}=0
\end{equation}

Finally, the equation for the $z$ coordinate becomes
\begin{equation}
\ddot{z}-\frac{1}{A}\frac{\partial A}{\partial \eta }\dot{z}\dot{\eta
}+A\frac{\partial A}{\partial z}{v_{T}}^{2}=0
\end{equation}

where the constant $v_{T}$ is defined by ${v_{T}}^{2}\equiv {v_{x}}^{2}+{v_{y}}^{2}$.
These equations of motion are generally not solvable by analytical
means. However, we will\ \ introduce an idealized model that allows
analytical treatment. The model is\ \ motivated by the observation
that, as shown in Appendix \ref{XRef-AppendixSection-9723255}, large
scale perturbations to the gravitational potential stay more or
less constant until the universe ceases to be matter dominated.

The idealized model that will be used from now on asserts a time-independent
gravitational potential with $\Phi ( \eta ) =\mathrm{const} = 1$
in the metric of eq. (\ref{XRef-Equation-918173048}).\ \ In this
case, the metric is conformally static, with
\begin{equation}
A( z) =1+2k( z-z_{0}) =A_{0}+2 k z%
\label{XRef-Equation-92018513}
\end{equation}

Notice that this is an unrealistic approximation at all but the
largest scales. At large, super-Hubble scales, still being a very
crude approximation, it is nevertheless a useful idealization that\ \ grants
us the luxury of analytical treatment and the ability of exploring
important characteristics of super-Hubble perturbations and their
effect on the CMB temperature anisotropy spectrum. As we will see
later on in this section, the temperature perturbation arizing from
a super-Hubble metric perturbation is roughly proportional to the
gain or loss in gravitational potential between the time of emission
of a CMB photon and the time of its observation. This implies that,
even if in a more realistic case, the super-Hubble perturbation
to the gravitational potential decays slightly at late times, this
decay is uniform at super-Hubble scales. Therefore, we would expect
the directional distribution of the temperature perturbation to
remain the same, modulo a time-dependent factor. Disregarding this
time-dependent factor, which is what we are doing, makes parameter
estimation imprecise, but does not change the qualitative signature
of the super-Hubble perturbation in the CMB spectrum.

Given this approximation, the equations of motion for a massless
particle now take the form
\begin{gather}
A \dot{\eta }=\mu =\mathrm{const}
\\\ddot{z}+A( z)  \frac{dA}{dz}{v_{T}}^{2}=0
\\A^{-1}\dot{x}= v_{x}, A^{-1}\dot{y}= v_{y}
\end{gather}

The four-velocity $u^{\mu }\equiv {\dot{x}}^{\mu }$ is a null vector,
which implies
\begin{equation}
A^{-2}{\dot{z}}^{2}+{v_{x}}^{2}+{v_{y}}^{2}={\dot{\eta }}^{2}%
\label{XRef-Equation-91912956}
\end{equation}

 Let us define the constant $A_{0}\equiv A( 0) =1-2 k z_{0}$ and
fix the integration constant $\mu $ by setting $\mu =A_{0}$\footnote{If
we were making this computation in a physical spacetime, we would
at this point relate $\dot{\eta }$ to the energy of the photon.
However, we must leave this aside, since we are computing the geodesic
in an unphysical spacetime. We therefore choose the boundary conditions
that are the most convenient.}. Furthermore, let us fix the affine
parameter $\lambda $ by demanding $z( 0) =0$. This gives $\dot{\eta
}( 0) =1$. We notice that impact angles and photon velocity are
conformally invariant quantities, and we are therefore free to compute
them in any conformal frame. Therefore, the physical velocity of
the photon in the frame of the observer is ${\hat{u}}^{i}=A^{-1}{\dot{x}}^{i}/\dot{\eta
}={\dot{x}}^{i}/A_{0}$, and eq. (\ref{XRef-Equation-91912956}) simply
states that $|\hat{u}|=1$. We therefore find that the integration
constants $v_{x}$ and $v_{y}$ are the velocities of the photon in
the frame of the observer at the instance of observation. Defining
$\theta $ as the observed angle of the photon, as measured from
the direction of positive $z$, we have:
\begin{equation}
{v_{T}}^{2}={\sin }^{2}\theta , \cos  \theta  = -{\hat{u}}^{z}=\dot{z}(
0) /A_{0}
\end{equation}

The boundary conditions are now defined that allow us to find unique
solutions to the equations of motion. Treating the case of vertical
motion with $\theta =0$ or $\theta =\pi $ separately, we obtain
the following solutions to the equation of motion, expressed in
terms of the observation angle $\theta $:
\begin{gather}
z( \lambda ) =\begin{cases}
\frac{A_{0}}{2k}\left( \frac{1}{ \sin  \theta }\sin ( \theta -2k
\sin  \theta  \lambda ) -1\right)  & 0<\theta <\pi  \\
- \lambda  \cos  \theta  & \theta =0,\pi  \\
\end{cases}
\\\eta ( \lambda ) =\left\{ \begin{array}{ll}
 \eta _{0}( \theta ) -\frac{1}{2k}\ln ( \tan ( \frac{1}{2}\left(
\theta -2k \sin  \theta  \lambda \right) ) )  & 0<\theta <\pi  \\
 -\frac{1}{2k \cos  \theta }\ln  \left( 1-\frac{2k\ \ \ \cos  \theta
}{A_{0}}\lambda \right)  & \theta =0,\pi
\end{array}\right.
\end{gather}

where $\eta _{0}( \theta ) \equiv \frac{1}{2k}\ln ( \tan ( \theta
/2) ) $. For simplicity, we have set $\eta =0$ at the time of observation.
We can now easily express the solution in terms of the conformal
time $\eta $. We get
\begin{equation}
z( \theta ,\eta ) =\begin{cases}
\frac{A_{0}}{2k}\left( \frac{1}{ \sin  \theta }\frac{1}{\cosh (
2k( \eta _{0}-\eta ) ) }-1\right)  & 0<\theta <\pi  \\
\frac{A_{0}}{2k}\left( e^{-2k \eta  \cos  \theta  }-1\right)   &
\theta =0,\pi  \\
\end{cases}%
\label{XRef-Equation-920144015}
\end{equation}

Eq. (\ref{XRef-Equation-920144015}) is parametrized by quantities
that are identitical in the two conformal frames, so it is a parametrization
that also is valid in the physical frame. It allows us to compute
the $z$ coordinate of a photon observed at an angle $\theta $ at
any time $\eta $ in its history. Next, we will use this to compute
the\ \ effect of a super-Hubble metric perturbation on the temperature
distribution of the CMB.

\subsection{CMB temperature distribution as measured by a static
observer in the CMB rest frame}

In the preceding section, we solved the null-geodesic equations
in the background of a super-Hubble metric pertubation, asserting,
for simplicity, that the\ \ pertubation is static in comoving coordinates:
\begin{equation}
{\mathrm{ds}}^{2}=a^{2}( \eta ) \left( -A( z) {\mathrm{d\eta }}^{2}+{A(
z) }^{-1}\left( {\mathrm{dx}}^{2}+{\mathrm{dy}}^{2}+{\mathrm{dz}}^{2}\right)
\right) %
\label{XRef-Equation-9219737}
\end{equation}

where $A( z) \equiv 1+2k( z-z_{0}) $. In this metric, the null geodesic
equation for the time coordinate $\eta $ takes the form ${\dot{p}}_{0}=0$,
where $p_{0}\equiv \partial L/\partial \dot{\eta }=-a^{2}A \dot{\eta
}$. This\ \ implies that $p_{0}$ is a constant of motion, which
we can relate to the energy of the photon. The photon energy as
observed by a static observer with 4-velocity $u^{\mu }=({(-g_{00})}^{-1/2},0,0,0)$
is
\[
E=-p_{\mu }u^{\mu }=-p_{0} {\left( -g_{00}\right) }^{-1/2}=-\frac{p_{0}
}{a\sqrt{A( z) }}
\]

For a photon with observed frequency $\nu $, $E=h \nu $. We therefore
have
\[
a( \eta ) \sqrt{A( z) }\nu =-p_{0}/h=\mathrm{const},
\]

a generalization of both the gravitational redshift formula of a
static universe and the cosmological redshift formula of a homogeneous,
expanding universe. We will assume that CMB photons were emitted
at time $\eta _{*}$ at a uniform temperature $T_{*}$. Since the
peak frequency of a black-body spectrum is proportional to the temperature
of the radiation, we can use this relationship to relate the temperature
at the time of emission to the observed temperature $T( \theta )
$ of CMB photons being observed at angle $\theta $ by a static observer
at the origin:
\begin{equation}
T( \theta ) \sqrt{A_{0}}=T_{*}a( \eta _{*}) \sqrt{A( z( \theta ,\eta
_{*}) ) },%
\label{XRef-Equation-9219119}
\end{equation}

where $A( z( \theta ,\eta _{*}) ) $ defines the gravitational field
at the time and place of emission, and $z( \theta ,\eta _{*}) $
is given by eq. (\ref{XRef-Equation-920144015}). Let $T_{0}$ be
the average CMB temperature at present, and let $\Theta ( \theta
) $ to be the temperature pertubation defined by $T( \theta ) )=T_{0}(
1+\Theta ( \theta ) ) $. Since $T_{0}/T_{*}=a( \eta _{*}) $, the
redshift relationship gives the following expression for the temperature
perturbation $\Theta ( \theta ) $ in terms of the gravitational
potential at the time of emission for a photon observed at angle
$\theta $:
\begin{equation}
\Theta ( \theta ) =\sqrt{\frac{A( z( \theta ,\eta _{*}) ) }{A_{0}}}-1
\end{equation}

With $A( z) =1+2k( z-z_{0}) $ and $z( \theta ,\eta ) $ given by
eq. (\ref{XRef-Equation-920144015}), the expression for the temperature
perturbation $\Theta ( \theta ) $ simplifies to
\begin{equation}
\Theta ( \theta ) =\sqrt{\frac{2\chi _{*}}{{\chi _{*}}^{2}+1}}\frac{1}{\sqrt{1+\frac{\left(
{\chi _{*}}^{2}-1\right) }{\left( {\chi _{*}}^{2}+1\right) }\cos
\theta }}-1,%
\label{XRef-Equation-92565921}
\end{equation}

where the constant $\chi _{*}$ is defined by $2k \eta _{*}=\ln
\chi _{*}$. $k$ is positive while $\eta _{*}<0$. $|\eta _{*}|$ is
of the order of the ${H_{0}}^{-1}$, the inverse of the current Hubble
parameter value. $k$ is the spatial derivative of a super-Hubble
perturbation to the potential $\Phi $. Therefore, using the definition
of a super-Hubble perturbation, we must have $|k \eta _{*}|<<1$.
Hence, $\chi _{*}$ is slightly less than 1. It is natural then to
introduce a small, positive, dimensionless parameter $\Phi _{*}\equiv
(1-\chi _{*})/2$ that will be useful for perturbative expansion
of the temperature $\Theta ( \theta ) $. Expanding $\Phi _{*}$ in
terms of $(k \eta _{*})$, we have, to lowest order in $(k \eta _{*})$:
\[
\Phi _{*}=-\left( k \eta _{*}\right) +\mathcal{O}( {\left( k \eta
_{*}\right) }^{2}) ,
\]

$\Phi _{*}$ can be interpreted as the loss in gravitational potential
since the time of emission for photons observed at angle $\theta
=0$. We will refer to $\Phi _{*}$ as the {\itshape potential anisotropy.
}Then, to second order in $\Phi _{*}$, $\Theta ( \theta ) $ is
\begin{equation}
\Theta ( \theta ) =\Phi _{*} \cos  \theta \ \ -{\Phi _{*}}^{2} \left(
1-\left( 1+\frac{3}{2} \cos  \theta \right) \cos  \theta \right)
+\mathcal{O}( {\Phi _{*}}^{3})
\end{equation}

The first-order term $\Phi _{*}\cos  \theta $ equals the temperature
perturbation expected from photons following the null geodesics
of a homogeneous FLRW universe in\ \ eq. (\ref{XRef-Equation-9219119}).
Hence, higher-order terms stem from deviations from geodesics in
a homogeneous space.\ \

In this section, we assumed an observer at rest in the CMB rest
frame. In reality, the frame of observation is moving relative to
the CMB rest frame.\ \ Our motion relative to the CMB rest frame
has been measured to about 370 km/s \cite{Hinshaw:2008kr}. In the
next section, we will transform the result of eq. (\ref{XRef-Equation-92565921})
to a moving frame of observation.

\subsection{CMB temperature distribution as measured by a moving
observer}

Let $ \mathcal{O}$ and $\mathcal{O}^{\prime }$ be two inertial observers,
$ \mathcal{O}$ at rest relative to the CMB, while $\mathcal{O}^{\prime
}$ moves with constant velocity $\text{\boldmath $v$}=v \hat{v}$,
where $\hat{v}$ is the unit vector in the direction of movement
and $v$ is the velocity relative to the CMB rest frame. Given that
$ \mathcal{O}$ observes a black-body distribution of CMB photons
at temperature $T( \hat{p}) $, with $\hat{p}$ being the direction
vector of the observed photon, observer $\mathcal{O}^{\prime }$
will observe a Doppler-shifted black-body spectrum with temperature
\cite{Peebles-1968}
\begin{equation}
T^{\prime }( {\hat{p}}^{\prime },\text{\boldmath $v$}) =T( \hat{p})
\frac{\sqrt{1-{\text{\boldmath $v$}}^{2}}}{\left( 1+v {\hat{p}}^{\prime
}\cdot \hat{v}\right) }%
\label{XRef-Equation-105223017}
\end{equation}

${\hat{p}}^{\prime }$ is the direction vector of the photon as observed
by observer $\mathcal{O}^{\prime }$, and ${\hat{p}}^{\prime }\cdot
\hat{v}$ is given by the aberration formula \cite{Einstein-1905a,Peebles-1968}
:
\begin{equation}
{\hat{p}}^{\prime }\cdot \hat{v}=\frac{\hat{p}\cdot \hat{v}-v}{\left(
1-v \hat{p}\cdot \hat{v}\right) }%
\label{XRef-Equation-105224125}
\end{equation}

Next, let us define two constants $\gamma \equiv \sqrt{\frac{2\chi
_{*}}{{\chi _{*}}^{2}+1}}$ and $\beta \equiv \frac{(1-{\chi _{*}}^{2})}{(1+{\chi
_{*}}^{2})}$. This allows us to write the temperature perturbation
of eq. (\ref{XRef-Equation-92565921}) as $\Theta ( \theta ) =\frac{\gamma
}{\sqrt{1-\beta  \cos  \theta }}-1=\frac{\gamma }{\sqrt{1+\beta
\hat{p}\cdot \hat{u}}}$, where $\hat{p}$ is the direction vector
of the photon, and $\hat{u}$ is the unit vector along the direction
of the gradient $\partial _{i}\Phi $ of the super-Hubble perturbation,
which we hereafter will call the anisotropy axis.

Now, let us use eq. (\ref{XRef-Equation-105223017}) to transform
the temperature distribution of eq. (\ref{XRef-Equation-92565921})
to the frame of observation:\ \ \ \
\begin{equation}
T^{\prime }( {\hat{p}}^{\prime },\text{\boldmath $v$}) =\frac{\gamma
}{\sqrt{1+\beta  \hat{p}\cdot \hat{u}}}T_{d}( {\hat{p}}^{\prime
},\text{\boldmath $v$}) ,%
\label{XRef-Equation-10661648}
\end{equation}

Here, we introduced the dipole distribution $T_{d}( {\hat{p}}^{\prime
},\text{\boldmath $v$}) \equiv T_{0}\sqrt{1-{v}^{2}}/(1+v {\hat{p}}^{\prime
}\cdot \hat{v}) $. Let us first compute $\hat{p}\cdot \hat{u}$ in
the frame of observation. Expanding the unit vectors $\hat{p}$ and
$\hat{u}$ in the CMB rest frame in spherical coordinates as $\hat{p}\equiv
-(\sin  \theta _{p} \cos  \phi _{p},\ \ \sin  \theta _{p} \sin
\phi _{p}, \cos  \theta _{p})$ and $\hat{u}\equiv (\sin  \theta
_{u} \cos  \phi _{u},\ \ \sin  \theta _{u} \sin  \phi _{u}, \cos
\theta _{u})$, we get $\hat{p}\cdot \hat{u}=-\sin  \theta _{p} \sin
\theta _{u} \cos ( \phi _{p}-\phi _{u}) -\cos  \theta _{p} \cos
\theta _{u}$. We can then transform to the frame of observation
by using the aberration formula of eq. (\ref{XRef-Equation-105224125})
and the fact that longitudinal coordinates are invariant under a
Lorentz boost along $\hat{v}$. We get, using $\theta \equiv \theta
_{p}^{\prime }$, $\phi \equiv \phi _{p}^{\prime }$, $\phi _{u}=\phi
_{u}^{\prime }=0$ and $\xi \equiv \cos  {\theta _{u}}^{\prime }
$:
\begin{equation}
f( \theta ,\phi ;\xi ) \equiv \hat{p}\cdot \hat{u}=-\frac{1}{\left(
1-v \cos  \theta \right) \left( 1-\xi  v\ \ \right) }\left( \left(
1-v^{2}\right) \sqrt{1-\xi ^{2}}\sin  \theta \ \ \cos  \phi \ \ +\left(
\xi -v\right) \left( \cos  \theta -v\right) \right)
\end{equation}

$(\theta ,\phi )$ are now angular coordinates in the frame of observation,
and $\xi $ is the dot product between the anisotropy axis and the
direction of motion in the frame of observation: $\xi \equiv {\hat{u}}^{\prime
}\cdot \hat{v}$.

If we assume a completely homogeneous temperature distribution in
the CMB rest frame, the distribution observed in the moving frame
of observation will, according to eq. (\ref{XRef-Equation-105223017}),
be the Doppler-shifted dipole distribution $T_{d}( {\hat{p}}^{\prime
}, \text{\boldmath $v$}) $.\ \ If we average this distribution over
the sphere, we get the average temperature in the frame of observation:
\[
T_{0}^{\prime }\equiv \frac{1}{4\pi }\int {\mathrm{d\Omega }}^{\prime
}T_{d}( {\hat{p}}^{\prime }, \text{\boldmath $v$}) =\sqrt{1-v^{2}}\frac{\operatorname{arctanh}(
v) }{v}T_{0}=\left( 1-\frac{v^{2}}{6}+\mathcal{O}( v^{3}) \right)
T_{0}
\]

We notice that the average temperature in the frame of observation
is slightly reduced compared to the static frame. The effect is
small, though, and proceeding, we will use $T_{0}^{\prime }=T_{0}$,
discarding terms higher than first order in velocity.

\section{The effect of a super-Hubble perturbation on the CMB Power
Spectrum}

Having computed the effect of a super-Hubble perturbation to the
gravitational potential on the observed CMB temperature distribution,
the next challenge is to elicit its effect on the CMB temperature
anisotropy power spectrum. The standard way of computing this power
spectrum is to first remove the dipole induced by the observatory's
peculiar motion through the CMB sky \cite{WMAP5-Explanatory-supplement},
then decompose the residual temperature distribution into multipoles,
and finally compute the covariance matrix of the multipole coefficients.
With a homogeneous background geometry, this covariance matrix becomes
diagonal. With an inhomogenous background geometry, however, the
covariance matrix obtains nodiagonal entries. Therefore, these nodiagonal
entries in the\ \ temperature anisotropy covariance matrix represent
the signature of the inhomogeneity of the background geometry.

\subsection{Factoring out the Doppler dipole}

Before computing the CMB power spectrum, the CMB dipole must be
removed from the temperature data \cite{WMAP5-Explanatory-supplement}.
Thereafter, the power spectrum may be computed from the residual
temperatures. Our moving observer $\mathcal{O}^{\prime }$ asserts
a homogeneous background, so we will therefore fit the temperature
distribution of eq. (\ref{XRef-Equation-10661648}) to that of a
Doppler-shifted uniform temperature distribution. The procedure
is lengthy, but straight forward. We will therefore describe it
briefly. Let us vary both the velocity and the average temperature
by writing the fitted velocity as $v^{{\prime\prime}}\equiv v+\delta
v$ and the fitted temperature average as $T_{0}^{{\prime\prime}}\equiv
T_{0}( 1+{\Delta T}_{0}) $. Defining the temperature residual as
$\delta T\equiv T_{d}( {\hat{p}}^{\prime },{\text{\boldmath $v$}}^{{\prime\prime}})
-T^{\prime }( {\hat{p}}^{\prime },\text{\boldmath $v$}) $, we find
the best fit dipole by minimizing the averaged squared error $\int
{\mathrm{d\Omega }}^{\prime }{\delta T}^{2}$with respect to variations
in velocity $v^{{\prime\prime}}$ and temperature $T_{0}^{{\prime\prime}}$.
The values of $\delta v$ and ${\Delta T}_{0}$ that provide the best
fit dipole distribution are
\begin{equation}
\delta v= \Phi _{*}( 1+\Phi _{*})  \xi -v \Phi _{*}( 1-\xi ^{2})
, {\Delta T}_{0}=-\frac{1}{2}{\Phi _{*}}^{2}
\end{equation}

Next, we will compute the temperature residual that remains after
the Doppler dipole has been removed from the temperature distribution.
The observed CMB temperature distribution of an assumed homogeneous
background metric has two factors; the best fit dipole distribution
and the observed residual distribution:
\[
T^{{\prime\prime}}( {\hat{p}}^{\prime },{\text{\boldmath $v$}}^{{\prime\prime}})
=T_{d}( {\hat{p}}^{\prime },{\text{\boldmath $v$}}^{{\prime\prime}})
\left( 1+\Theta ^{{\prime\prime}}( {\hat{p}}^{\prime }) \right)
\]

Here, $\Theta ^{{\prime\prime}}$ is the residual temperature perturbation
that remains after factoring out the best fit dipole.

Similarly, the expected temperature distribution of the inhomogeneous
background of eq. (\ref{XRef-Equation-9219737}) has three factors;
the anisotropy factor given by eq. (\ref{XRef-Equation-92565921})
, the dipole distribution and some arbitrary residual temperature
distribution $T\equiv 1+\Theta $ given a priori:
\[
T^{\prime }( {\hat{p}}^{\prime },\text{\boldmath $v$}) =\frac{\gamma
}{\sqrt{1+\beta  f( {\hat{p}}^{\prime },\xi ) }}T_{d}( {\hat{p}}^{\prime
},\text{\boldmath $v$}) \left( 1+\Theta ( {\hat{p}}^{\prime }) \right)
\]

Using eq. (\ref{XRef-Equation-10661648}), we may then relate the
observed temperature pertubation $\Theta ^{{\prime\prime}}$ to the
given perturbation $\Theta $:\ \
\begin{equation}
\Theta ^{{\prime\prime}}( {\hat{p}}^{\prime }) =\frac{\gamma }{\sqrt{1+\beta
f( {\hat{p}}^{\prime },\xi ) }}\frac{T_{d}( {\hat{p}}^{\prime },\text{\boldmath
$v$}) }{T_{d}( {\hat{p}}^{\prime },{\text{\boldmath $v$}}^{{\prime\prime}})
}\left( 1+\Theta ( {\hat{p}}^{\prime }) \right) -1%
\label{XRef-Equation-10618117}
\end{equation}

Eq. (\ref{XRef-Equation-10618117}) will be used later in the final
computation of the CMB anisotropy power spectrum.

\subsection{CMB anisotropy spectrum}

The temperature distribution $\Theta ^{{\prime\prime}}( \hat{p})
$ can be expanded in terms of spherical harmonics $Y_{l}^{m}( \hat{p})
$ and multipole coefficients $a_{\mathrm{lm}}$ as $\Theta ^{{\prime\prime}}(
\hat{p}) =\operatorname*{\Sigma }\limits_{l}\operatorname*{\Sigma
}\limits_{m=-l}^{l}a_{\mathrm{lm}}Y_{l}^{m}( \hat{p}) $. This equation
can be inverted, expressing the\ \ multipole coefficients\ \ $a_{\mathrm{lm}}$
in terms of the field $\Theta ^{{\prime\prime}}$:
\begin{equation}
a_{\mathit{lm}}=\int \mathrm{d\Omega } {Y_{l}^{m}( \hat{p}) }^{*}\Theta
^{{\prime\prime}}( \hat{p}) %
\label{XRef-Equation-106184154}
\end{equation}

Let us define the CMB covariance function as the expectation value
of the square of the multipole coefficients:
\begin{equation}
\mathcal{C}_{l,k}^{m,n}\equiv \left\langle  a_{\mathit{kn}}a_{\mathit{lm}}^{*}\right\rangle
\end{equation}

For a homogeneous background geometry, $\mathcal{C}_{l,k}^{m,n}$
can be expressed in terms of the $C_{l}$ of the conventional CMB
anisotropy power spectrum, in which case it takes the simple, diagonal
form
\[
\mathcal{C}_{l,k}^{m,n}=C_{l}\delta _{l,k}\delta _{m,n}
\]

Now, let us compute $\mathcal{C}_{l,k}^{m,n}$ for the inhomogeneous
background of eq. (\ref{XRef-Equation-9219737}). Using eq. (\ref{XRef-Equation-106184154}),
we get\ \
\[
\mathcal{C}_{l,k}^{m,n}=\int {\mathrm{d\Omega }}^{\prime } \int
\mathrm{d\Omega } Y_{k}^{n}( {\hat{p}}^{\prime }) {Y_{l}^{m}( \hat{p})
}^{*}<\Theta ^{\prime \prime }( {\hat{p}}^{\prime })  \Theta ^{\prime
\prime }( \hat{p}) >
\]

Now, $\Theta ^{{\prime\prime}}$ is given by eq. (\ref{XRef-Equation-10618117}).
Define $\Delta \Theta $ as
\begin{equation}
\Delta \Theta \equiv \frac{\gamma }{\sqrt{1+\beta  f( {\hat{p}}^{\prime
},\xi ) }}\frac{T_{d}( {\hat{p}}^{\prime },\text{\boldmath $v$})
}{T_{d}( {\hat{p}}^{\prime },{\text{\boldmath $v$}}^{{\prime\prime}})
}-1%
\label{XRef-Equation-10619541}
\end{equation}

Now, using eq. (\ref{XRef-Equation-10618117}) , the observed temperature
perturbation $\Theta ^{{\prime\prime}}$ can be expressed in terms
of $\Delta \Theta $ as follows:
\[
\Theta ^{{\prime\prime}}( {\hat{p}}^{\prime }) =\left( 1+\Delta
\Theta \right) \left( 1+\Theta ( {\hat{p}}^{\prime }) \right) -1
\]

This allows us to express $\mathcal{C}_{l,k}^{m,n}$ in terms of
$\Delta$$\Theta$:
\begin{equation}
\mathcal{C}_{l,k}^{m,n}=\int {\mathrm{d\Omega }}^{\prime }Y_{k}^{n}(
{\hat{p}}^{\prime })  \Delta \Theta {\left( {\hat{p}}^{\prime }\right)
}^{*}\int \mathrm{d\Omega } {Y_{l}^{m}( \hat{p}) }^{*}\Delta \Theta
( \hat{p}) +\int {\mathrm{d\Omega }}^{\prime } Y_{k}^{n}( {\hat{p}}^{\prime
}) \left( 1+\Delta \Theta ( {\hat{p}}^{\prime }) \right) \int \mathrm{d\Omega
} Y_{l}^{m}{\left( \hat{p}\right) }^{*}\left( 1+\Delta \Theta (
\hat{p}) \right) <{\Theta ( {\hat{p}}^{\prime }) }^{*}\Theta ( \hat{p})
>%
\label{XRef-Equation-106191058}
\end{equation}

The expectation value $<{\Theta ( {\hat{p}}^{\prime }) }^{*}\Theta
( \hat{p}) >$ can be expressed in terms of the Fourier-transformed
temperature distribution $\Theta ( k,\text{\boldmath $k$}\cdot \hat{p})
$, the matter overdensity $\delta ( k) $ and the matter power spectrum
$P( k) $ (see eq. (8.65) of \cite{Dodelson-2003} ):
\[
<{\overline{\Theta }( {\hat{p}}^{\prime }) }^{*}\overline{\Theta
}( \hat{p}) >=\int \frac{d^{3}k}{{\left( 2\pi \right) }^{3}}P( k)
\frac{{\Theta ( k,\text{\boldmath $k$}\cdot {\hat{p}}^{\prime })
}^{*}}{\delta ( k) } \frac{\Theta ( k,\text{\boldmath $k$}\cdot
\hat{p}) }{\delta ( k) }
\]

The first term on the right-hand side of eq. (\ref{XRef-Equation-106191058})
is second order in the perturbation variables $\Phi _{*}$ and $v$,
so we will discard it in the following. This gives
\[
\mathcal{C}_{l,k}^{m,n}=\int {\mathrm{d\Omega }}^{\prime } Y_{k}^{n}(
{\hat{p}}^{\prime }) \left( 1+\Delta \Theta ( {\hat{p}}^{\prime
}) \right) \int \mathrm{d\Omega } Y_{l}^{m}{\left( \hat{p}\right)
}^{*}\left( 1+\Delta \Theta ( \hat{p}) \right) \int \frac{d^{3}k}{{\left(
2\pi \right) }^{3}}P( k) \frac{{\Theta ( k,\text{\boldmath $k$}\cdot
{\hat{p}}^{\prime }) }^{*}}{\delta ( k) } \frac{\Theta ( k,\text{\boldmath
$k$}\cdot \hat{p}) }{\delta ( k) }
\]

First, we may expand $\Theta ( k,\text{\boldmath $k$}\cdot \hat{p})
$ in terms of multipoles $\Theta _{l}( k) $:
\[
\Theta ( k,\text{\boldmath $k$}\cdot \hat{p}) ={\operatorname*{\Sigma
}\limits_{l}( -i) }^{l}\left( 2l+1\right) \mathcal{P}_{l}( \hat{k}\cdot
\hat{p}) \Theta _{l}( k)
\]

Furthermore, the Legendre polynomials $\mathcal{P}_{l}$ can be expanded
in terms of spherical harmonics:
\[
\mathcal{P}_{l}( \hat{k}\cdot \hat{p}) =\frac{4\pi }{2l+1}\operatorname*{\Sigma
}\limits_{m=-l}^{l}Y_{l}^{m}( \hat{p}) {Y_{l}^{m}( \hat{k}) }^{*}
\]

Now, define the parameter $\tau \equiv \Phi _{*}\sqrt{1-\xi ^{2}}$.
The covariance function $\mathcal{C}_{l,k}^{m,n}$ then takes the
form
\begin{equation}
\mathcal{C}_{l,k}^{m,n}=C_{l}\delta _{k,l}\delta _{m,n}+\tau (
C_{k}+ C_{l}) \ \ K_{l,k}^{m,n}%
\label{XRef-Equation-106192547}
\end{equation}

where $C_{l}$ is defined by
\begin{equation}
C_{l}\equiv \frac{2}{\pi }\int d k k^{2}\frac{P( k) }{{\delta (
k) }^{2}}{\left| \Theta _{l}( k) \right| }^{2}%
\label{XRef-Equation-10619353}
\end{equation}

and the coefficients $K_{l,k}^{m,n}$ are defined by the integral
\begin{equation}
K_{l,k}^{m,n}\equiv \tau ^{-1}\int \mathrm{d\Omega } \Delta \Theta
( \hat{p})  {Y_{l}^{m}( \hat{p}) }^{*}Y_{k }^{n}( \hat{p}) %
\label{XRef-Equation-10619516}
\end{equation}

We see that $C_{l}$ represents the variance of the multipole coefficients
$a_{\mathrm{lm}}$, and its definition in eq. (\ref{XRef-Equation-10619353})
is indeed the conventional formula for computing the\ \ variance
of the multipole coefficients $a_{\mathrm{lm}}$ for a homogeneous
background, see p. 242 of \cite{Dodelson-2003}. Notice that when
the expression for $\mathcal{C}_{l,k}^{m,n}$ in eq. (\ref{XRef-Equation-106192547})
was computed,\ \ terms of order $\tau ^{2}$ were discarded.

Eq. (\ref{XRef-Equation-106192547}) is the general form of the covariance
function $\mathcal{C}_{l,k}^{m,n}$. If the coefficients $K_{l,k}^{m,n}$
are nonzero, the covariance function is nondiagonal. However, we
have yet to compute the coefficients $K_{l,k}^{m,n}$ and prove that
there are indeed nondiagonal entries in the covariance function.

\subsection{Nondiagonal entries of the covariance function}

Next, we will compute the nondiagonal coefficients $K_{l,k}^{m,n}$
of the covariance function $\mathcal{C}_{l,k}^{m,n}$. The coefficients
are given by eq. (\ref{XRef-Equation-10619516}). Before we are able
to compute the integral, we must compute $\Delta \Theta $ from its
definition in eq. (\ref{XRef-Equation-10619541}). To first order
in the perturbation variables $\Phi _{*}$ and $v$, we get\ \ \
\begin{equation}
\Delta \Theta =\Phi _{*}\sqrt{1-\xi ^{2}}\sin  \theta  \cos  \phi
\label{XRef-Equation-101955217}
\end{equation}

Spherical harmonics may be written in terms of the associated Legendre
functions $P_{l}^{m}$ (see p. 670 of \cite{Riley-Hobson-Bence-1998}):
\[
Y_{l}^{m}( \theta ,\phi ) =\sqrt{\frac{\left( 2l+1\right) }{4\pi
}\frac{\left( l-m\right) !}{\left( l+m\right) !}}P_{l}^{m}( \cos
\theta ) e^{i m \phi }
\]

$K_{l,k}^{m,n}$ can then be expressed as follows, defining the angular
variable $\mu \equiv \cos  \theta $:\ \ \
\[
K_{l,k}^{m,n}=\sqrt{\frac{\left( 2l+1\right) }{4\pi }\frac{\left(
l-m\right) !}{\left( l+m\right) !}}\sqrt{\frac{\left( 2k+1\right)
}{4\pi }\frac{\left( k-n\right) !}{\left( k+n\right) !}}\operatorname*{\int
}\limits_{-1}^{1}\mathrm{d\mu } \sqrt{1-\mu ^{2}}P_{l}^{m}( \mu
) P_{k}^{n}( \mu )  \operatorname*{\int }\limits_{0}^{2\pi }\mathrm{d\phi
} e^{i \left( n-m\right)  \phi }\cos  \phi
\]

Now, we may compute the integral over the longitudinal variable
$\phi $:
\[
\operatorname*{\int }\limits_{0}^{2\pi }\mathrm{d\phi } e^{i \left(
n-m\right)  \phi }\cos  \phi  =\pi ( \delta _{m,n+1}+\delta _{m+1,n})
\]

Furthermore, we may use the following recursion relationship for
the associated Legendre functions \cite{Abramovitz-Stegun,Bell}:
\[
\sqrt{1-\mu ^{2}}P_{l}^{m-1}( \mu ) =\frac{1}{\left( 2l+1\right)
}\left( P_{l+1}^{m}( \mu ) -P_{l-1}^{m}( \mu ) \right)
\]

as well as the orthogonality relationship (see p. 66 of \cite{Bell}
):
\[
\operatorname*{\int }\limits_{-1}^{1}\mathrm{d\mu } P_{l}^{m}( \mu
) P_{k}^{m}( \mu ) =\frac{2}{2l+1}\frac{\left( l+m\right) !}{\left(
l-m\right) !}\delta _{l,k}
\]

By introducing the coefficients
\begin{equation}
c_{l}^{m}\equiv \begin{cases}
\sqrt{\frac{l+m}{2l+1}} & -l\leq m\leq l \\
0 & |m|>l \\
\end{cases}
\end{equation}

we obtain the following expression for the\ \ coefficients $K_{l,k}^{m,n}$:
\begin{equation}
K_{l,k}^{m,n}=\frac{1}{2}\left( \left( c_{l}^{m}c_{k}^{m}\delta
_{m,n+1}-c_{l}^{-m}c_{k}^{-m}\delta _{m,n-1}\right) \delta _{l,k+1}
+\left( c_{l}^{n}c_{k}^{n}\delta _{m+1,n}-c_{l}^{-n}c_{k}^{-n}\delta
_{m-1,n}\right) \delta _{l+1,k}\right) %
\label{XRef-Equation-106214438}
\end{equation}

The coefficients $K_{l,k}^{m,n}$ have the following symmetries:
\begin{equation}
K_{l,k}^{m,n}=K_{k,l}^{n,m}
\end{equation}

and
\begin{equation}
K_{l,k}^{m,n}=-K_{l,k}^{-m,-n}%
\label{XRef-Equation-101822363}
\end{equation}

\subsection{Estimating the potential anisotropy vector}

The perturbation parameter $\Phi _{*}$ measures the loss in gravitational
potential for photons arriving along the anisotropy axis ${\hat{u}}^{\prime
}$. Therefore, $\Phi _{*}\xi $ measures the gain or loss in gravitational
potential for photons arriving along the dipole axis at $\theta
=0$, while\ \ $\Phi _{*}\sqrt{1-\xi ^{2}}$ measures the gain or
loss in gravitational potential for photons arriving from directions
orthogonal to the dipole axis, at $\theta =\pi /2$.\ \

We refer to the vector $\Phi _{*}{\hat{u}}^{\prime }$ as the {\itshape
potential anisotropy vector}. $\Phi _{*}\sqrt{1-\xi ^{2}}$ is then
the transverse component of the potential anisotropy vector,\ \ which
is the component orthogonal to the dipole axis, while $\Phi _{*}\xi
$ is the longitudianal component of the potential anisotropy vector,
which is the component along the dipole axis. Define $\lambda \equiv
\Phi _{*}\xi $. The parameter $\tau \equiv \Phi _{*}\sqrt{1-\xi
^{2}}$ has already been introduced. We will refer to $\tau $ as
the {\itshape transverse potential anisotropy} and $\lambda $ as
the {\itshape longitudinal potential anisotropy}. In order to estimate
the direction of the the anisotropy vector, we also need to estimate
the longitude of the anisotropy axis in galactic coordinates.Alternatively,
by noticing that the temperature anisotropy $\Delta \Theta =\tau
\sin  \theta  \cos  \phi $ has extrema at $\theta =\pi /2, \phi
=\pm \pi $, we could estimate the direction of the potential anisostropy
based on an estimate of the direction of maximal temperature anisotropy
perpendicular to the CMB dipole axis plus estimates of the parameters
$\tau $ and $\lambda $. If $\hat{w}$ is the observed direction of
maximal temperature asymmetry, we have
\begin{equation}
\Phi _{*}{\hat{u}}^{\prime }=\tau  \hat{w}+\lambda  \hat{v}%
\label{XRef-Equation-102163428}
\end{equation}

We see from eq. (\ref{XRef-Equation-106214438}) that to lowest order
in the perturbation parameters, we are only able to estimate the
transverse potential anisotropy $\tau $. In order to estimate the
longitudinal potential anisotropy, the covariance function\ \ $\mathcal{C}_{l,k}^{m,n}$
must be evolved to higher order in the perturbation variables. In
the scope of this paper, we will restrict ourselves to lowest order
estimates, so we will therefore only be able to estimate the transverse
potential anisotropy.

From eq. (\ref{XRef-Equation-106214438}) we obtain four sets of
estimators for $\tau $, one for each of the four terms. The first
and third terms give the same estimator, and so do the second and
fourth terms. The remaining two estimators can be identified by
using the symmetry relation of eq. (\ref{XRef-Equation-101822363}).\ \ The
basic estimators can then be written
\begin{equation}
\hat{\tau }=\frac{2\mathcal{C}_{l,l-1}^{m,m-1}}{\left( C_{l}+C_{l-1}\right)
c_{l}^{m}c_{l-1}^{m}}, m=-\left( l-2\right) ,...,\left( l-1\right)
\end{equation}

This basic estimator can be used for estimation of the transverse
potential anisotropy $\tau $ from multipole coefficients $a_{\mathrm{lm}}$
derived from CMB observations. For each multipole index $l$, we
may sample $\tau $\ \ $2(l-1)$ times.\ \ If we choose a set of $N$
multipoles ranging from $l=2$ to $l=N+1$, we will therefore be able
to retrieve $\operatorname*{\Sigma }\limits_{l=2}^{N+1}2(l-1)=N(
N+1) $ independent samples of\ \ $\tau $, thus reducing the uncertainty
of the estimate considerably.

\subsection{Hemispherical power asymmetry}

Finally, let us have a glance at the hemispherical power asymmetry
that would result from the temperature anisotropy $\Delta \Theta
=\tau  \sin  \theta  \cos  \phi $ of\ \ eq. (\ref{XRef-Equation-101955217}).
Designating $(\theta =\frac{\pi }{2},\phi =\pi )$ as the north pole,
the northern hemisphere is parametrized by $\pi /2<\phi <3\pi /2$.
We would therefore expect a maximal temperature difference between
the two hemispheres to be $\sim 2\tau  \Theta $, yielding a power
difference between the two hemispheres of the order of $\tau  C_{l}$.
Furthermore, given that the north pole is at $\theta =\pi /2$, the
axis of maximal power asymmetry should be perpendicular to the CMB
dipole axis. Given a CMB dipole at $(l,b)=(264\mbox{}^{\circ},48\mbox{}^{\circ})$
\cite{Hinshaw:2008kr} and a best fit power asymmetry axis at $(l,b)=(225\mbox{}^{\circ},-27\mbox{}^{\circ})$\cite{Eriksen:2007pc},
the angle separating\ \ the two axes is 83$ \mbox{}^{\circ}$. Table
\ref{XRef-TableTitle-1020103642} shows the angle of separation for
each of the 5 different analyses made by Eriksen et. al \cite{Eriksen:2007pc}.
In each case, the angle of separation between the CMB dipole axis
and the axis of maximal power asymmetry lies between 83$ \mbox{}^{\circ}$
and 96$ \mbox{}^{\circ}$, consistent with the prediction by our
model.\ \ \

We notice that eq. (\ref{XRef-Equation-102163428}) can be applied
to estimate the direction of the potential anisotropy from estimates
of the power asymmetry axis $\hat{w}$,\ \ given estimates of\ \ parameters
$\tau $ and $\lambda $.
\begin{table}
\caption{Separation angle between the dipole axis and the axis of
maximal power asymmetry for different results of Eriksen et. al.
\cite{Eriksen:2007pc}. {\itshape a) small sky cut applied. b) large
sky cut applied}} \label{XRef-TableTitle-1020103642}
\begin{ruledtabular}

\begin{tabular}{lcc}
Data & (l,b) & Separation angle\\
\hline
${\mathrm{ILC}}^{a}$ & (225$ \mbox{}^{\circ}$, -27$ \mbox{}^{\circ}$)
& 83$ \mbox{}^{\circ}$\\
${\mathrm{ILC}}^{b}$ & (208$ \mbox{}^{\circ}$, -27$ \mbox{}^{\circ}$)
& 90$ \mbox{}^{\circ}$\\
Q-band & (222$ \mbox{}^{\circ}$, -35$ \mbox{}^{\circ}$) & 91$ \mbox{}^{\circ}$\\
V-band & (205$ \mbox{}^{\circ}$, -19$ \mbox{}^{\circ}$) & 85$ \mbox{}^{\circ}$\\
W-band & (204$ \mbox{}^{\circ}$, -31$ \mbox{}^{\circ}$) & 96$ \mbox{}^{\circ}$
\end{tabular}
\end{ruledtabular}
\end{table}

An interesting side note is that the anomalous CMB Cold Spot \cite{Vielva:2003et,Cruz:2004ce,Cruz:2006fy}
is located in the same region of the sky as the north pole of the
power anistotropy axis: The cold spot is located at $(l,b)=(209\mbox{}^{\circ},
-57\mbox{}^{\circ})$. Given that there is a CMB power asymmetry
with a minimal power in this particular region and assuming this
power asymmetry can be explained by the presence of a super-Hubble
inhomogeneity in the gravitational field, the cold spot will be
less extreme, because the expectation value for the CMB temperature
would be lower in this particular region. Consequently, the cold
spot should be less anomalous in this case, as already noted by
Eriksen et. al. \cite{Eriksen:2007pc}.

\section{Conclusions}

Prompted by the puzzling evidence of an hemispherical power asymmetry
in the CMB \cite{Eriksen:2003db}, this paper started with two simple
questions: 1):\ \ How could a large-scale inhomogeneity in the gravitational
field of super-Hubble size - larger than the observable universe
- be observed? And 2): Could an hemispherical\ \ power asymmetry
in the CMB be caused by a super-Hubble scale inhomogeneity in the
gravitational field enclosing the present Hubble volume? Posing
these questions, we made no assumptions about the origin of this
perturbation or any physical mechanism that might have caused it.
Our goal was simply to determine what effect the presence of such
a perturbation to the gravitational potential would have on the
observed CMB temperature and to elicit its observational signature
in the CMB spectrum.

In order to tackle these questions by analytical means, we introduced
an idealized model in which large-scale perturbations to the gravitational
potential stay constant in comoving coordinates. This model is an
unrealistic approximation at all but the largest scales. At large,
super-Hubble scales, still being a very crude approximation\footnote{We
notice that in reality, large scale perturbations to the potential
decay uniformly about 30\% in a flat $\Lambda$CDM universe with
$\Omega _{m}=0.3$}, it is nevertheless a useful idealization that
grants us the luxury of analytical treatment and the ability of
exploring important characteristics of super-Hubble perturbations
and their effect on the CMB temperature anisotropy spectrum. We
find that the temperature perturbation arizing from a super-Hubble
metric perturbation is roughly proportional to the loss in gravitational
potential between the time of emission of a CMB photon and the time
of its observation. Therefore, even if in a more realistic model,
the super-Hubble perturbation to the gravitational potential decays
slightly at late times, this decay is uniform at super-Hubble scales.
Therefore, given a more realistic model, we still expect the directional
distribution of the temperature perturbation to remain the same
as in our idealized model, modulo a time-dependent factor. Disregarding
this time-dependent factor, which is what we are doing, makes parameter
estimation imprecise, but does not change the qualitative signature
of the super-Hubble perturbation in the CMB spectrum.\ \ \

Using our idealized model, we were able to solve the null-geodesic
equations of motion and determine the resulting CMB temperature
distribution in the CMB rest frame. Transforming the resulting temperature
distribution to a moving frame of observation and removing the best
fit Doppler dipole, we obtained a residual temperature distribution.
By expanding the residual temperature distribution into multipoles,
we obtained our main result: the covariance function of the multipole
coefficients.

The covariance function contains, in addition to the $C_{l}$ entries
of the conventional CMB temperature anisostropy power spectrum,
non-diagonal entries.We find that the $C_{l}$ entries of the conventional
anisotropy power spectrum are insensitive to the strength of the
potential anisotropy. Thus, the non-diagonal entries, which are
not present in the case of a homogeneous background geometry, constitute
the main signature of a large-scale inhomogeneity in the background
geometry of the universe. This anwers the first of our two initial
questions.

Regarding the second of our initial questions, an inhomogeneity
in the gravitational potential of super-Hubble size would yield
a power asymmetry in the CMB with maximal asymmetry at an angle
of 90$ \mbox{}^{\circ}$ to the CMB dipole axis.\ \ The power asymmetry
that was observed appears to be at an angle that lies between 83$
\mbox{}^{\circ}$ and 96$ \mbox{}^{\circ}$ to the CMB dipole axis,
which is consistent with the prediction of our model. This is suggestive
of a simple explanation for the CMB power asymmetry, because it
implies that the location of the observed power asymmetry in the
CMB sky can be accounted for by a large-scale inhomogeneity in the
gravitational field enclosing the present Hubble volume. At this
point, we will not claim that our model can completely account for
the CMB power asymmetry. More work remains. In particular, it still
remains to be seen whether this model is also able to\ \ account
for the power asymmetry in quantitative terms.

\appendix

\section{Super-Hubble solution to the first order perturbation equations
for $\Lambda$CDM}
\label{XRef-AppendixSection-9723255}

Let us start by taking a look at the evolution equation for the
temperature perturbation $\Theta \equiv \Delta T/T$. We use conformal
Newtonian gauge and apply the notation and exposition of Dodelson
(see ch. 7 of \cite{Dodelson-2003}).\ \ $\Theta $ is a function,
not only of conformal time $\eta $ and position $x$, but also of
photon direction, ${\hat{p}}^{i}$. The first order perturbation
equation for $\Theta ( \eta ,x,\hat{p}) $ is
\begin{equation}
\frac{\partial \Theta }{\partial \eta }+\frac{\partial \Phi }{\partial
\eta }+{\hat{p}}^{i}\partial _{i}\Theta +{\hat{p}}^{i}\partial _{i}\Psi
=n_{e}\sigma _{T}a( \Theta _{0}-\Theta +\hat{p}\cdot {\overset{\rightarrow
}{v}}_{b}) %
\label{XRef-Equation-9121215}
\end{equation}

where $n_{e}$ is the number density of free electrons, ${\overset{\rightarrow
}{v}}_{b}$ is the baryon velocity and $\sigma _{T}$ is the Thomson
cross section for electron-photon scattering. $\Theta _{0}$ is the
temperature monopole: $\Theta _{0}( \eta ,x) \equiv \frac{1}{4\pi
}\int d\Omega  \Theta ( \hat{\eta ,x,p}) $. Seeking super-Hubble
solutions to eq. (\ref{XRef-Equation-9121215}), we may discard the
terms containing spatial derivatives. Integrating the equation by
$\frac{1}{4\pi }\int d\Omega $, eq. (\ref{XRef-Equation-9121215})
simplifies, leaving an equation for the monopole $\Theta _{0}$ and
the potential $\Phi $:\ \
\begin{equation}
\frac{\partial \Theta _{0}}{\partial \eta }+\frac{\partial \Phi
}{\partial \eta }=0
\end{equation}

Here, we have used that $\int d\Omega  \hat{p}\cdot \overset{\rightarrow
}{u}=0$ for any direction-independent vector $\overset{\rightarrow
}{u}$, which implies that $\int d\Omega \hat{p}\cdot {\overset{\rightarrow
}{v}}_{b}=0$.

For super-Hubble perturbations we may therefore discard baryon and
photon interactions, which occur at much smaller scales. At super-Hubble
scales, perturbations to the dark matter and baryon distributions
can be treated as a common distribution of matter. Similarly, at
super-Hubble scales, perturbations to the distributions of photons
and neutrinos can be treated the same, as a common distribution
of relativistic radiation.

The two Einstein equations determining the evolution of the metric
perturbations take the form
\begin{gather}
 3\mathcal{H} \frac{\partial \Phi }{\partial \eta }-\partial ^{2}\Phi
-3\mathcal{H}^{2}\Psi  =4\pi  G a^{2}( \rho _{m}^{\left( 0\right)
}\delta +4\rho _{r}^{\left( 0\right) }\Theta _{0})
\\\partial ^{2}\left( \Phi +\Psi \right) =0%
\label{XRef-Equation-91214717}
\end{gather}

Here, $\delta \equiv {\delta \rho }_{m}/\rho _{m}^{(0)}$ is the
matter overdensity, while $\Theta _{0}$ is the monopole moment of
the radiation distribution. $\rho _{m}^{(0)}$ is the zero-order
matter density, which takes the form $\rho _{m}^{(0)}=\rho _{\mathrm{cr}}\Omega
_{m}/a^{3}$, where $\rho _{\mathrm{cr}}$ is the critical density
today and $\Omega _{m}$ is the present matter ratio. $\rho _{r}^{(0)}$
is the zero-order radiation density, which takes the form $\rho
_{r}^{(0)}=\rho _{\mathrm{cr}}\Omega _{r}/a^{4}$.

The equation for the dark matter density perturbation is
\begin{equation}
\frac{\partial \delta }{\partial \eta }+3\frac{\partial \Phi }{\partial
\eta }+\partial _{i}v^{i}=0,
\end{equation}

while the equation for the dark matter velocity $v^{i}$ is
\begin{equation}
\frac{\partial v^{i}}{\partial \eta }+\mathcal{H} v^{i}+\partial
_{i}\Psi  =0
\end{equation}

Eq. (\ref{XRef-Equation-91214717}) is easily solved by setting $\Psi
=-\Phi $. Again, as we seek super-Hubble solutions to the equations,
we discard terms with spatial derivatives. In that case, the velocity
equations for matter decouple from the other equations, and we are
left with the three equations for the three perturbation variables
$\Phi $, $\Theta _{0}$ and $\delta $. Furthermore, as the equations
now only contain time derivatives of these variables, we may separate
each perturbation variable $X( \eta ,x) $ into a time-independent
factor and a time-dependent factor. The time-independent factors
can be determined by applying initial conditions.\ \ We apply adiabatic
initial conditions, which are $\Phi ( \eta _{i},x) =\Phi _{0}( x)
$,\ \ $\Theta _{0}( \eta _{i},x) =\frac{1}{2}\Phi _{0}( x) $ and
$\delta ( \eta _{i},x) =\frac{3}{2}\Phi _{0}( x) $, where $\eta
_{i}$ is an arbitrarily chosen initial time. $\Phi _{0}( x) $ is
an arbitrary perturbation in the potential satisfying $|\partial
_{i}\Phi _{0}|<< \mathcal{H}$ and $\partial ^{2}\Phi _{0}<< \mathcal{H}^{2}$.
Hence, the time-indepdent factor of each of the three perturbation
variables are $\Phi _{0}( x) $.\ \ We will therefore rewrite the
variables as follows: $\Theta _{0}( \eta ,x) \rightarrow \Theta
_{0}( \eta ) \Phi _{0}( x) $, $\delta ( \eta ,x) \rightarrow \delta
( \eta ) \Phi _{0}( x) $ and $\Phi ( \eta ,x) \rightarrow \Phi (
\eta ) \Phi _{0}( x) $. This allows us to factor out the spatial
dependence of the equations entirely, and we are left with the following
ordinary differential equations, using overdots to denote differentiation
with respect to $\eta $:
\begin{gather}
{\overset{\cdot }{\Theta }}_{0}+\dot{\Phi }=0%
\label{XRef-Equation-9123431}
\\3\mathcal{H} \dot{\Phi }+3\mathcal{H}^{2}\Phi =4\pi  G a^{2}(
\rho _{m}^{\left( 0\right) }\delta +4\rho _{r}^{\left( 0\right)
}\Theta _{0}) %
\label{XRef-Equation-9123853}
\\\dot{\delta }+3\dot{\Phi }=0,%
\label{XRef-Equation-912355}
\end{gather}

Applying adiabatic initial conditions $\Phi ( \eta _{i}) =2\Theta
_{0}( \eta _{i}) $ and $\delta ( \eta _{i}) =3\Theta _{0}( \eta
_{i}) $ at an initial time $\eta _{i}$, allows us to integrate eqs.
(\ref{XRef-Equation-9123431})\ \ and (\ref{XRef-Equation-912355}).
We get
\[
\Theta _{0}( \eta ) =\frac{\delta ( \eta ) }{3}, \delta ( \eta )
=\frac{9}{2}-3\Phi ( \eta )
\]

Changing to a new temporal variable $x\equiv \ln  a,$ eq. (\ref{XRef-Equation-9123853})
can now be written\ \
\begin{equation}
\frac{\mathrm{d\Phi }}{dx}+\Phi  =\frac{1}{2}\frac{\left( \Omega
_{m}e^{-3x}+\frac{4}{3}\Omega _{r}e^{-4x}\right) }{ \Omega _{\Lambda
}+\Omega _{m}e^{-3x}+\Omega _{r}e^{-4x}}\left( \frac{9}{2}-3\Phi
\right) %
\label{XRef-Equation-926240}
\end{equation}

There are no known analytical solutions to eq. (\ref{XRef-Equation-926240}),
but we may find approximate solutions that cover the entire history
of the universe. First, let us seek an approximate solution that
covers the radiation and matter-dominated eras. In this case $a^{3}
<<1$. Introducing a new variable $y=\frac{\Omega _{m}}{\Omega _{r}}e^{x}=\frac{\Omega
_{m}}{\Omega _{r}}a$\ \ \ \ allows us to rewrite eq. (\ref{XRef-Equation-926240})
as follows:
\begin{equation}
y \frac{d\Phi }{dy}+\left( 1+\frac{1}{2}\frac{3y+4}{y+1}\right)
\Phi  =\frac{3}{4}\frac{\left( 3y+4\right) }{\left( y+1\right) }%
\label{XRef-Equation-9263251}
\end{equation}

Eq. (\ref{XRef-Equation-9263251}) has the following solution that
satisfies the initial condition that requires $\Phi ( y) \rightarrow
1$ as $y\rightarrow 0$:
\begin{equation}
\Phi =\frac{1}{10 y^{3}} \left( -16-8 y+2 y^{2}+9 y^{3}+16\sqrt{1+y}\right)
\end{equation}

We see that $\Phi $ has an almost constant value of $1$ throughout
the radiation era, then drops slightly to approach $9/10$ in the
matter-dominated era.

Next, let us seek an approximate solution to eq. (\ref{XRef-Equation-9263251})
that is valid in the matter-dominated era and later. In eq. (\ref{XRef-Equation-9263251}),
we can now discard the radition terms. Defining a new variable $z\equiv
\frac{\Omega _{\Lambda }}{\Omega _{m}}e^{3x}$, we get the equation\ \ \ \ \ \ \ \
\[
z \frac{d\Phi }{dz}+\frac{2z +5 }{6 \left( z+1\right) }\Phi  =\frac{3}{4}\frac{1}{
z+1}
\]

Its solution can be expressed in terms of the hypergeometric function
${}_{2}F_{1}$:
\begin{equation}
\Phi =\frac{3 }{2}-\frac{3}{5}\ \ \left( 1+z\right) {}_{2}F_{1}
\left( 1,\frac{4}{3};\frac{11}{6};-z\right)
\end{equation}

This solution has the value $\Phi ( 0) =9/10$. At present times,
$\Phi \simeq 0.7$ for $\Omega _{m}=0.3$.\label{1}\label{Comment1}\label{potential-decay}

\end{document}